\documentstyle[12pt,epsfig]{article}
\begin{document}

\begin{titlepage}
\rightline{June 2011}
\vskip 2cm
\centerline{\Large \bf
Mirror \& hidden sector dark matter}
\vskip 0.5cm
\centerline{\Large \bf 
in the light of new CoGeNT data}

\vskip 2.2cm
\centerline{R. Foot\footnote{
E-mail address: rfoot@unimelb.edu.au}}

\vskip 0.7cm
\centerline{\it ARC Centre of Excellence for Particle Physics at the Terascale,}
\centerline{\it School of Physics, University of Melbourne,}
\centerline{\it Victoria 3010 Australia}
\vskip 2cm
\noindent
The CoGeNT collaboration has recently made available new 
data collected over a period of 15 months. In addition to 
more accurately measuring the spectrum of nuclear recoil candidate events they
have announced evidence for an annual modulation signal. 
We examine the implications of these new results
within the context of mirror/hidden sector dark matter models. We find that the new CoGeNT
data can be explained within this framework with parameter space consistent with 
the DAMA annual modulation signal, and the null results of the other experiments.
We also point out that the CoGeNT spectrum at low energies is observed to obey 
$dR/dE_R \propto 1/E_R^2$ which suggests that dark matter interacts via 
Rutherford scattering
rather than the more commonly assumed contact (four-fermion) interaction.

\end{titlepage}

The CoGeNT experiment operating in the Soudan Underground Laboratory has been searching for light
dark matter interactions with a low energy threshold P-type Point Contact 
germanium detector\cite{cogent,cogent2}.
They have obtained a low energy spectrum which is not readily explainable in terms of known
background sources, and is consistent with elastic scattering
of light dark matter particles\cite{foot1,foot2,hooper}.
Recently, 15 months of data has been analyzed\cite{cogent2} which greatly improves the
measurement of the low energy spectrum  
and appears to be annually modulated with a phase consistent with
dark matter expectations\cite{dm}. This further strengthens the
dark matter interpretation of the CoGeNT low energy spectrum.

The recent CoGeNT results also reinforce the long standing observation of the
DAMA collaboration of an annual modulation signal in their $NaI$ detector\cite{dama1,dama2}.  
The DAMA signal is extremely impressive
with statistical significance of $8.9$ sigma with phase and period in agreement with the dark matter
expectations to high accuracy.  Attempts to interpret the DAMA signal in terms of a hypothetical 
background have become more and more implausible, and there is ample reason to be confident that
DAMA and now CoGeNT have observed dark matter. 

The positive results of DAMA and CoGeNT together with the null results of very sensitive, but
higher threshold experiments such as CDMS\cite{cdmsge} and XENON100\cite{xenon100} suggest that dark matter is 
light ($\stackrel{<}{\sim} 30$ GeV).
Having dark matter light, has long been known to alleviate the tension between DAMA and the higher
threshold experiments\cite{footold} (see also ref.\cite{gelmini}). However, the sensitivity 
of the higher threshold
experiments has got to the point where there is now some tension between DAMA/CoGeNT
and e.g. XENON10. XENON100, CDMS/Si, CDMS/Ge when interpreted in terms of standard WIMPs 
even if they are light\cite{gelmini2}.
Another issue is that the allowed parameter regions for the DAMA and CoGeNT signals, although
close together do not significantly overlap in the standard WIMP framework\cite{hooper}.  
It turns out that both of these difficulties can
be resolved if dark matter is not only light, but assumed to be multi-component and self interacting.

A generic example of this\cite{foot2008} is dark matter from a hidden sector, which contains an 
unbroken $U(1)'$ gauge interaction
which is mixed with the standard $U(1)_Y$ via renormalizable kinetic mixing interaction:\cite{he}
\begin{eqnarray}
{\cal L}_{mix} = \frac{\epsilon'}{2\cos\theta_w} F^{\mu \nu} F'_{\mu \nu}
\label{kine}
\end{eqnarray}
where $F_{\mu \nu}$ is the standard $U(1)_Y$ gauge boson 
field strength tensor, and $F'_{\mu \nu}$ is the field strength tensor for the hidden sector $U(1)'$. 
This interaction enables hidden sector $U(1)'$ charged particles (of charge $Qe$) to couple to
ordinary photons with electric charge $Q\epsilon' e \equiv \epsilon e$.
We consider the case where the hidden sector contains two (or more) stable $U(1)'$ charged 
dark matter particles, 
$F_1$ and $F_2$
with masses $m_1$ and $m_2$ [$F_1$ and $F_2$ can be fermionic or bosonic]. 
Under the standard assumptions of a dark halo forming an
isothermal sphere the condition of hydrostatic
equilibrium relates the temperature of the particles to
the galactic rotational velocity, $v_{rot}$:
\begin{eqnarray}
T = {1 \over 2} \bar m v_{rot}^2
\end{eqnarray}
where $\bar m \equiv {n_{F_1} m_1 + n_{F_2} m_2 \over n_{F_1} + n_{F_2}}$
is the mean mass of the particles in the galactic halo.
We have assumed that the self interactions mediated by the unbroken $U(1)'$ gauge
interactions are sufficiently strong so that they thermalize the hidden sector 
particles, $F_1$ and $F_2$. The interaction length is typically much less than a parsec\cite{footz} and the
dark matter particles form a pressure-supported halo.
The dark matter particles are then described by a Maxwellian
distribution with $f_i(v) = exp(-E/T) = exp (-{1 \over 2} m_i v^2/T) = exp[-v^2/v_0^2 (i)]$
where
\begin{eqnarray}
v_0 (i) = v_{rot} \sqrt{{\bar m \over m_i}}\ . 
\label{sun}
\end{eqnarray}
With the assumptions that $m_2 >> m_1$ and that the abundance of $F_2$ is much less than $F_1$, we
have that $v_0^2 (F_2) \ll v_{rot}^2$. The narrow velocity dispersion (recall
$\sigma_v^2 = 3v_0^2/2$) can greatly reduce the rate
of dark matter interactions in higher threshold experiments such as XENON100 whilst still explaining the
signals in the lower threshold DAMA and CoGeNT experiments.  

While generic hidden sector models are interesting in their own right and have been
studied in some detail (see e.g. ref.\cite{feng}), I consider 
mirror dark matter as the best motivated example of such a multi-component
self-interacting theory.
Recall, mirror dark matter posits that the inferred dark matter in the Universe arises from
a hidden sector which is an exact copy of the standard model sector\cite{flv} (for a review
and more complete list of references see ref.\cite{review})\footnote{
Note that successful big bang nucleosynthesis and successful
large scale structure requires effectively asymmetric initial
conditions in the early Universe, $T' \ll T$ and $n_{b'}/n_b \approx 5$. 
See ref.\cite{some} for further discussions.}.
That is, 
a spectrum of dark matter particles of known masses are predicted: $e', H', He', O', Fe',...$ (with
$m_{e'} = m_e, m_{H'} = m_H,$ etc). 
The galactic halo is then presumed to be composed predominately of a spherically distributed 
self interacting mirror particle plasma comprising these particles\cite{sph}. 
Kinetic mixing of the $U(1)_Y$ and its mirror counterpart 
allows ordinary and mirror particles to interact with each other and can thereby
explain the direct detection experiments\cite{foot1,foot2,footold}. 
The simplest scenario involves kinetic mixing induced
elastic (Rutherford) scattering of the dominant mirror metal component, $A'$, off target nuclei.
[The $He'$ and $H'$ components are too light to give a signal above the DAMA/CoGeNT energy threshold]. 
Previous work\cite{foot1,foot2} (see also ref.\cite{footold,foot2008}) has shown that such 
elastic scattering can explain the 
normalization and energy dependence of the DAMA annual 
modulation amplitude and also the initial (56 day) CoGeNT spectrum 
consistently with the null results of the other experiments, 
and yields a measurement of $\epsilon \sqrt{\xi_{A'}}$ and $m_{A'}$:
\begin{eqnarray}
\epsilon \sqrt{\xi_{A'}} &\approx & (7 \pm 3)\times 10^{-10}, \nonumber \\
\frac{m_{A'}}{m_p} &\approx & 22\pm 8
\label{bla}
\end{eqnarray}
where $\xi_{A'} \equiv n_{A'}m_{A'}/(0.3 \ GeV/cm^3)$ is the halo mass fraction of the species
$A'$ and $m_p$ is the proton mass. 
The measured value of $m_{A'}/m_p$ is consistent with $A' \sim O'$, which
by analogy with the ordinary matter sector would be the naive expectation.
Taking a range for $\xi_{A'}$, $1 \stackrel{>}{\sim} \xi_{A'} \stackrel{>}{\sim} 10^{-2}$, suggests that 
$\epsilon$ could realistically range from $10^{-10}$ to $10^{-8}$.
Kinetic mixing in this range is consistent with laboratory and  
astrophysical constraints\cite{lab1} and has a number of 
fascinating applications\cite{lab1,review,mitra}. Early Universe cosmology, though,
prefers\cite{disf} $\epsilon \stackrel{<}{\sim} 10^{-9}$.

The interaction rate in experiments depends on the halo distribution function
and the interaction cross-section. The former is expected to be a Maxwellian distribution,
$f_i(v) = exp[-v^2/v_0^2(i)]$, with $v_0(i)$ depending on $\bar m$, as discussed above. In the mirror
dark matter case, $\bar m$ is expected to be around 1 GeV, but with significant uncertainties\cite{foot1,foot2}.
Generally it has been found\cite{foot1} that the dark matter detection experiments are relatively insensitive
to the precise value of $v_0 (A')$ (and hence $\bar m$) so long as $v_0^2 (A') \ll v_{rot}^2$.
Kinetic mixing induced elastic Rutherford scattering is particularly natural in mirror/hidden sector models
as it arises from the renormalizable interaction, Eq.(\ref{kine}). In the present study
we again assume that it is the dominant interaction mechanism coupling ordinary and dark matter.
The cross-section for a dark matter particle of charge
$\epsilon e$ 
to elastically scatter off an ordinary nucleus (presumed at rest with mass 
and atomic numbers $A,\ Z$) is given by\cite{footold}:\footnote{We employ natural units where
$\hbar = c = 1$.}
\begin{eqnarray}
{d\sigma \over dE_R} = {\lambda \over E_R^2 v^2}
\label{cs}
\end{eqnarray}
where 
\begin{eqnarray}
\lambda \equiv {2\pi \epsilon^2 Z^2 \alpha^2 \over m_A} F^2_A (qr_A)   \
\end{eqnarray}
and $F_A (qr_A)$ is the form factor which
takes into account the finite size of the nuclei. In the case where dark matter
particles also have finite size, as in the mirror dark matter case, a form factor for those particles 
also needs to 
be included. [For elastic scattering of mirror nuclei, $A'$,
of atomic number $Z'$ we must replace $\epsilon \to Z'\epsilon$ 
in the above cross-section formula].
A simple analytic expression for
the form factor, which we adopt in our numerical work, is the one
proposed by Helm\cite{helm,smith}.

The event rate is given by:
\begin{eqnarray}
{dR \over dE_R} = N_T n_{A'} 
\int^{\infty}_{|{\textbf{v}}| > v_{min}}
{d\sigma \over dE_R}
{f_{A'}({\textbf{v}},{\textbf{v}}_E) \over k} |{\textbf{v}}| d^3v 
\label{55}
\end{eqnarray}
where $N_T$ is the number of target nuclei per kg of detector and  
$n_{A'} = \rho_{dm} \xi_{A'}/m_{A'}$ is the number density of halo dark matter particles $A'$ at the Earth's
location (we take $\rho_{dm} = 0.3 \  GeV/cm^3$).
Here ${\bf{v}}$ is the velocity of the halo particles relative to the
Earth and ${\bf{v}}_E$ is the
velocity of the Earth relative to the galactic halo.
The integration limit, $v_{min}$, is given by the kinematic relation:
\begin{eqnarray}
v_{min} &=& \sqrt{ {(m_{A} + m_{A'})^2 E_R \over 2 m_{A} m^2_{A'} }}\ .
\label{v}
\end{eqnarray}
The halo distribution function in the reference frame of the Earth is given by, 
$f_{A'} ({\bf{v}},{\bf{v}}_E)/k = (\pi v_0^2[A'])^{-3/2} exp(-({\bf{v}}
+ {\bf{v}}_E)^2/v_0^2[A'])$. The integral, Eq.(\ref{55}), can easily be evaluated in terms
of error functions\cite{foot2008,smith} and numerically solved.

To compare with the measured event rate, we must include detector resolution effects 
and overall detection efficiency (when the latter is not already included in the
experimental results):
\begin{eqnarray}
{dR \over dE_R^m} = \epsilon_f (E_R^m) {1 \over \sqrt{2\pi}\sigma_{res} } 
\int {dR \over dE_R} e^{-(E_R - E_R^m)^2/2\sigma^2_{res}} dE_R 
\end{eqnarray}
where $E_R^m$ is the measured energy and $\sigma_{res}$ describes the resolution.
The measured energy is typically in keVee units (ionization/scintillation energy). For nuclear recoils
in the absence of any channeling, $keVee = keV/q$, where $q < 1$ is the 
relevant quenching factor. Channeled events, where target atoms travel down crystal
axis and planes, have $q \simeq 1$. In light of recent theoretical studies\cite{newstudy}, we assume
that the channeling fraction is negligible. It is of course still possible that
channeling could play some role, which could modify the favoured regions of 
parameter space somewhat.

For this study we consider two of the simplest examples of multi-component
dark matter models.  Following our earlier works\cite{foot1,foot2,foot2008, footold}
we consider mirror dark matter with a 
dominant mirror metal component, $A'$, of 
atomic number $Z'$. In this case the electric charge of the dark matter particle is $\epsilon Z' e$.
\footnote{
In our numerical work we allow 
$A', Z'$, to have non-integer values,
with $Z' = A'/2$. Since the realistic case will involve a spectrum of
elements, the effective mass can be non-integer.}
The quantity $v_0(A')$ is obtained from Eq.(\ref{sun}) with
$\bar m \simeq 1.1 $ GeV, which corresponds to a $He'$ dominated halo,
$Y_{He'} \simeq 0.9$, expected\cite{bbn} for $\epsilon \sim 10^{-9}$.
We also consider the more generic two component $F_1,\ F_2$ hidden sector dark matter model discussed above,
in which case $v_0$ is less constrained.

One can define a $\chi^2$ quantity and compare these theories with experiment.
We consider the reference point $v_{rot} = 240$ km/s which is
representative of recent measurements for the local rotational velocity\cite{reidrecent}.
The data we consider consists of (a) the CoGeNT energy spectrum: 31 bins of
width $\Delta E \simeq 0.05$ keVee given in the inset of figure 1 of ref.\cite{cogent2}. This spectrum
has already been corrected for efficiency and stripped of background components. 
(b) The DAMA annual modulation energy spectrum in the energy range $2 < E(keVee) < 8$.
We have taken into account systematic uncertainties in energy scale by minimizing 
$\chi^2$ over a $20\%$ variation in quenching factors, i.e.
$q_{Na} = 0.30 \pm 0.06, \ q_I = 0.09 \pm 0.02$ for DAMA and $q_{Ge} = 0.21 \pm 0.04$ 
for CoGeNT. 
The mirror dark matter candidate provides an excellent fit to the data,
with $\chi^2_{min}/d.o.f$ values of 23.1/29 for data set (a),
and 8.9/10 for data set (b).\footnote{
The low CoGeNT threshold of $0.45$ keVee potentially makes the experiment sensitive 
to the $e'$ component
via $e'$-electron scattering, which would be expected to lead
to a large rise in event rate at low energies\cite{footescattering}. 
The data is adequately fit by $A'-Ge$ elastic scattering, with no evidence for an extra $e'-e$ 
scattering contribution. This
suggests that the $e'$ halo component has a lower temperature than
the mirror nuclei component.  Such a scenario is possible due to the inefficient energy transfer between
the light $e'$ and much heavier mirror nuclei.}
Favoured regions in the 
$\epsilon\sqrt{\xi_{A'}}, m_{A'}$ plane can be obtained by evaluating contours corresponding to
$\chi^2 (\epsilon\sqrt{\xi_{A'}}, m_{A'}) = \chi^2_{min} + 9$ (roughly $99\%$ C.L. allowed region). 
In figure 1 we show the parameter regions favoured by
the data for the $v_{rot} = 240$ km/s reference point.
The favoured regions for the DAMA and CoGeNT signals are in as good an agreement 
as one might expect given the
systematic uncertainties which we have not considered including the
fiducial bulk volume uncertainty in CoGeNT of $\sim 10\%$ and variation of $v_{rot}$ within
its estimated $\sim 10\%$ uncertainty.

\vskip 0.2cm
\centerline{\epsfig{file=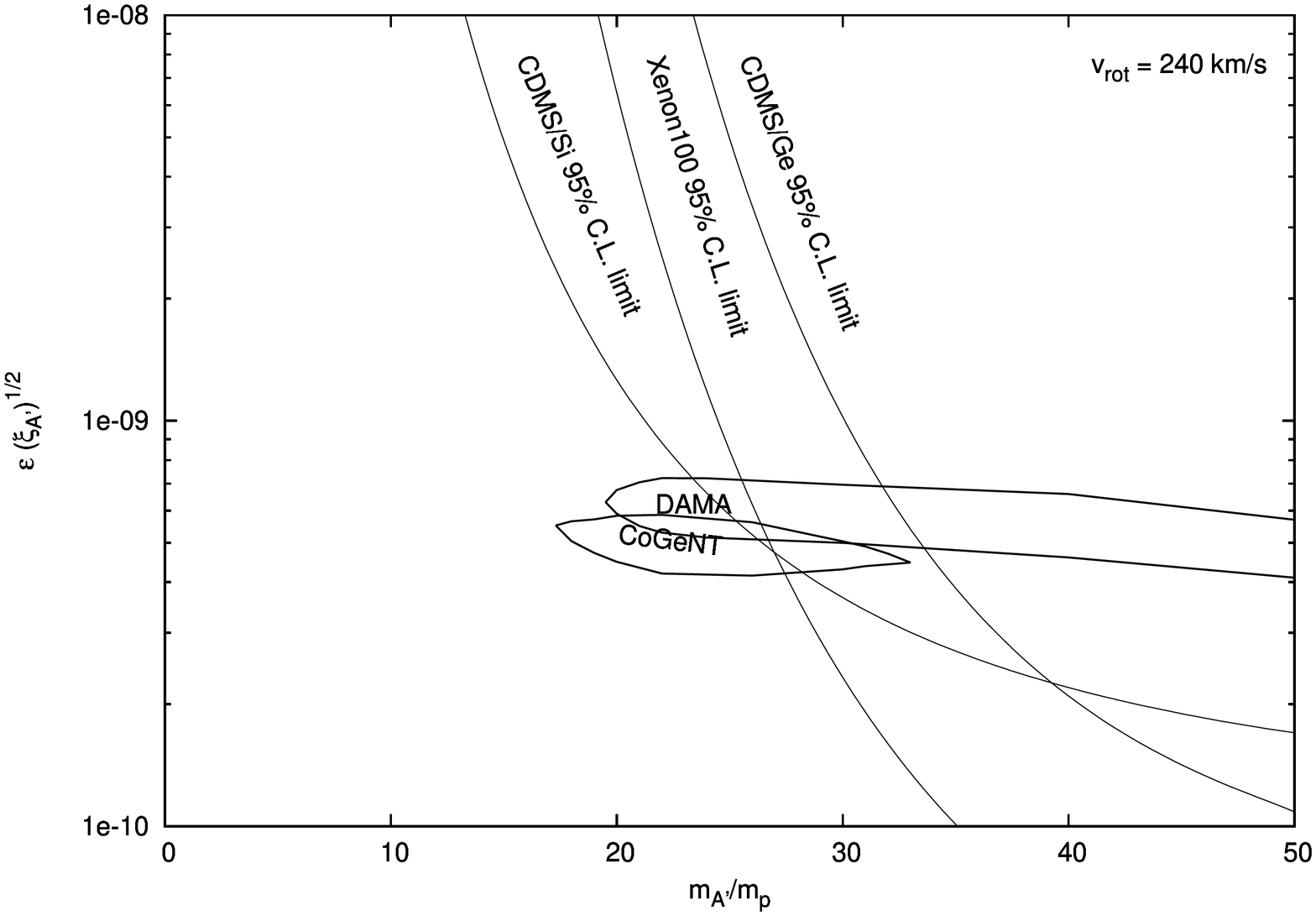,angle=270,width=15.0cm}}
\vskip 0.3cm
\noindent
{\small
Figure 1: CoGeNT and DAMA $99\%$ C.L. favoured parameter ($m_{A'}, \epsilon \sqrt{\xi_{A'}}$)
regions for the mirror dark matter model.  The reference point
$v_{rot} = 240$ km/s is assumed. Also shown are the $95\%$ exclusion curves
evaluated from the null results of XENON100, CDMS/Si and CDMS/Ge.
}

\vskip 0.8cm
\noindent
Also displayed in figure 1 is the $95\%$ exclusion limits evaluated for the CDMS/Si\cite{cdmssi},
CDMS/Ge\cite{cdmsge} and XENON100\cite{xenon100} experiments\footnote{
There are also lower threshold analysis by XENON10\cite{xenonlow} and CDMS\cite{cdmslow} 
collaborations. However when
systematic uncertainties are properly incorporated, neither analysis is capable of
excluding light dark matter explanations of the DAMA/CoGeNT signal\cite{collarlow}.}.
In computing these limits, we have conservatively
taken the energy thresholds of these experiments to be $20\%$ higher than the advertised values, to
allow for systematic uncertainties in energy calibration and quenching factor
\footnote{Within the mirror dark matter framework the higher threshold experiments such as CDMS/Ge
and XENON100 have an important role in probing the heavier $\sim Fe'$ component\cite{mm3}.}.
We also show in figure 2a,b, the predicted results for each 
data set for a particular parameter point near the global best fit, as well as a point
near the best fit for each data set considered separately.

\vskip -0.3cm
\centerline{\epsfig{file=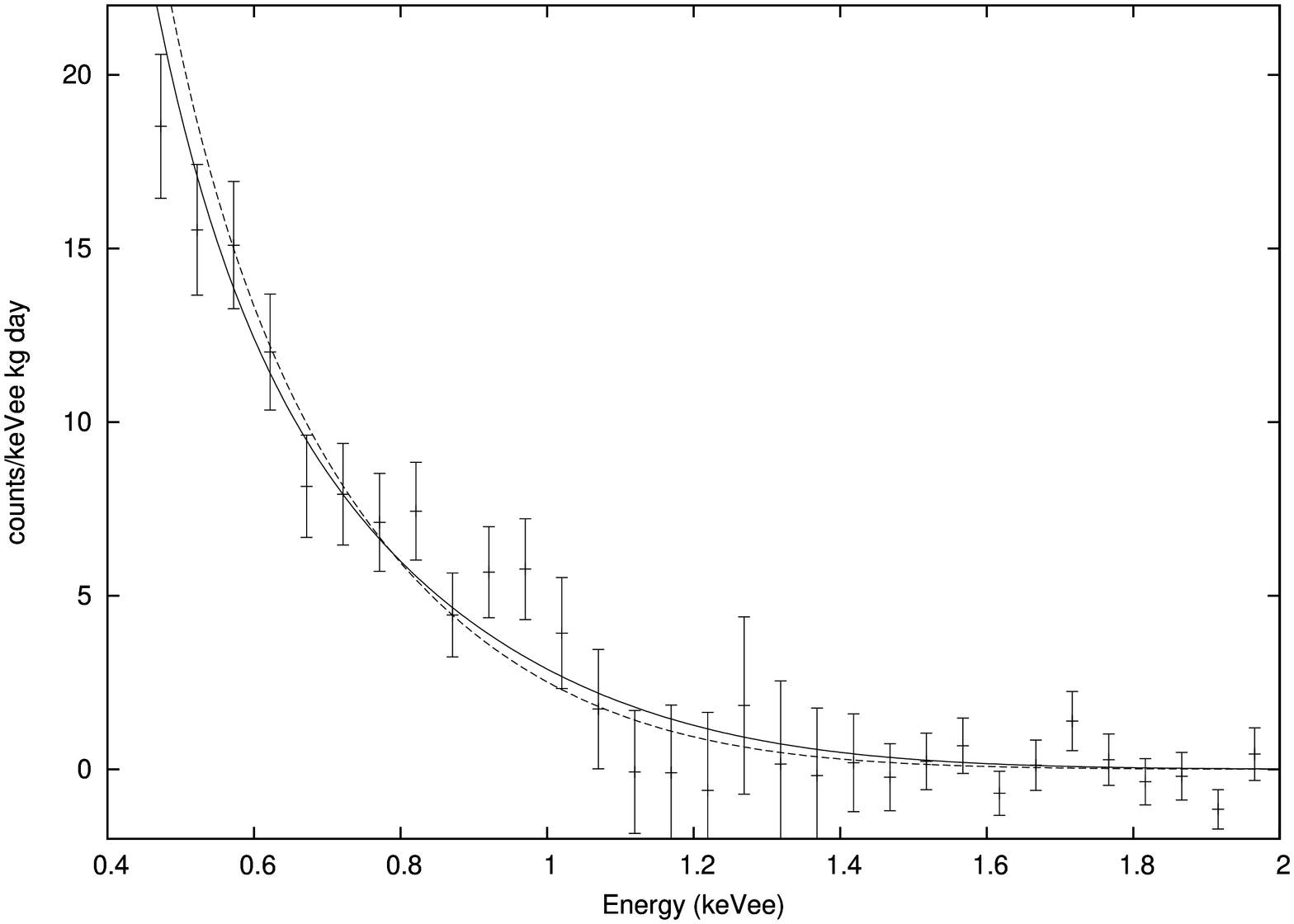,angle=270,width=13.4cm}}
\vskip 0.3cm
\noindent
{\small Figure 2a: Mirror dark matter versus the CoGeNT spectrum.
The solid line is for a point near the CoGeNT best 
fit [$m_{A'}/m_p = 26, \ \epsilon \sqrt{\xi_{A'}} = 5.2 \times 10^{-10}$]
while the dashed line is for a point near
the global best fit [$m_{A'}/m_p = 24, \ \epsilon \sqrt{\xi_{A'}} = 5.7\times 10^{-10}$].
}
\vskip 0.1cm

\centerline{\epsfig{file=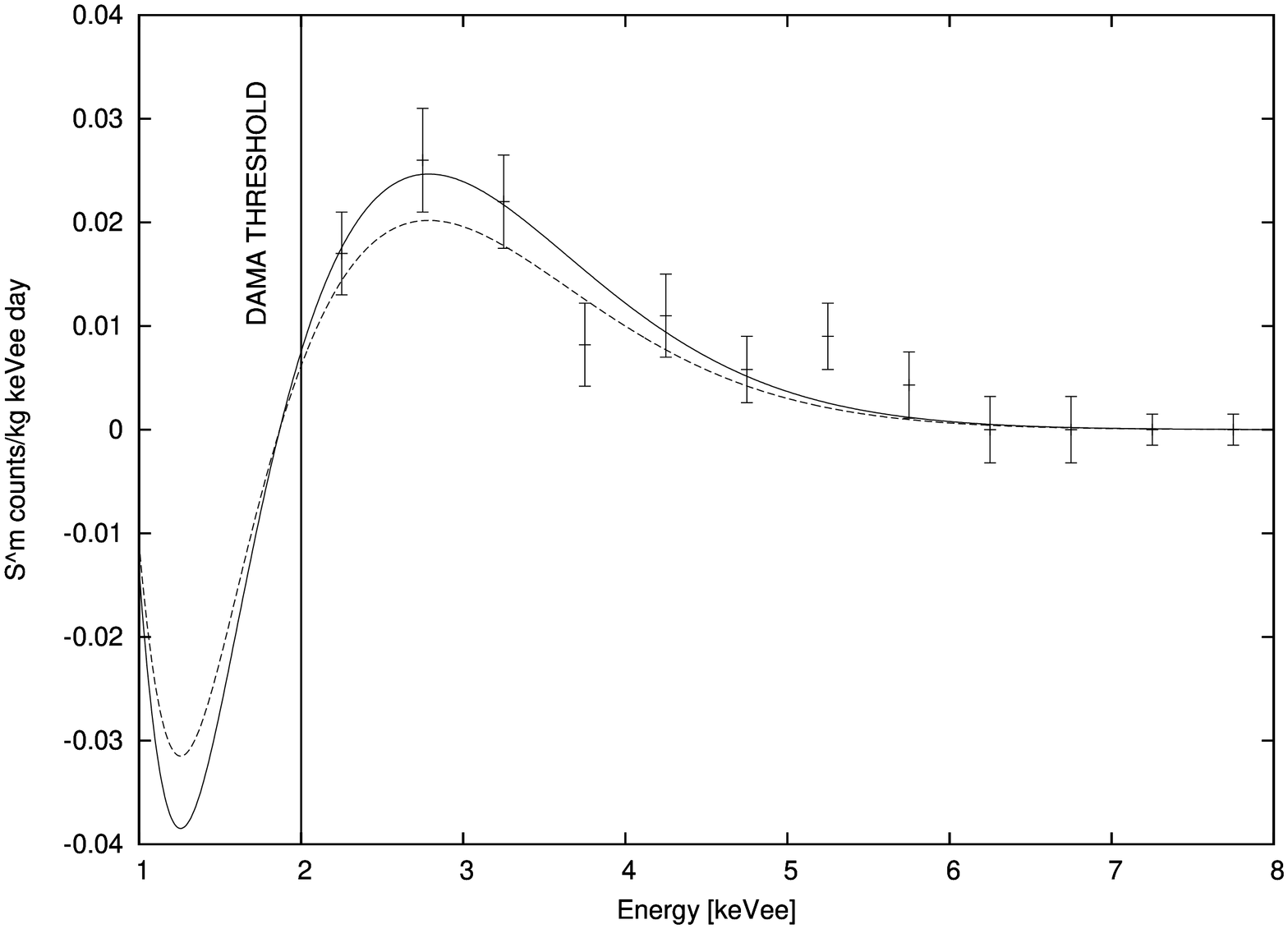,angle=270,width=13.4cm}}
\vskip 0.3cm
\noindent
{\small Figure 2b: Mirror dark matter versus the DAMA annual modulation spectrum.
The solid line is for a point near the DAMA best fit 
[$m_{A'}/m_p = 24, \ \epsilon \sqrt{\xi_{A'}} = 6.3\times 10^{-10}$] while the
dashed line is for the point near the global best fit considered in figure 2a
[$m_{A'}/m_p = 24, \ \epsilon \sqrt{\xi_{A'}} = 5.7\times 10^{-10}$]. 
}

\vskip 0.5cm

It is interesting to compare the 15 month CoGeNT favoured region, as shown in figure 1, 
with results for the same
model obtained with the initial 56 days of data\cite{foot1,foot2}.
The current favoured region is significantly reduced in size. CoGeNT data now
feature an upper limit on $m_{A'} \stackrel{<}{\sim} 30$ GeV, which is
also supported by the null results of XENON100 and CDMS/Si.

The CoGeNT collaboration report\cite{cogent2} evidence for an annual modulation signal in their data at
about $2.8 \sigma$ C.L. The amplitude of the modulation, averaged over $0.5 < E(keVee) < 3.0$,
is measured to be roughly
$A \approx 0.46 \pm 0.17$ cpd/kg/keVee. This assumes the amplitude and phase are set to theoretical
expectations, while a larger amplitude is preferred if the phase is left free.
For the theories offered here, we find that
the CoGeNT annual modulation amplitude
(averaged over $0.5 < E(keVee) < 3.0$) is typically around $\sim 0.12$ cpd/kg/keVee for the
parameter region near the global best fit, and does not get above $0.20$ for any parameter point
in the global $99\%$ C.L. favoured parameter region (for the reference point $v_{rot} = 240$ km/s). 
Thus we find an annual modulation somewhat below the CoGeNT central value.
This difference, though, is not currently,
statistically significant, representing only a 1.5-2 sigma downward fluctuation from the
central value measured in the $15\times 0.33$ month-kg data sample. Obviously
future data, especially the measurement of the energy spectrum of the annual modulation
amplitude, will be important tests of the theories considered here.

Similar results hold for the more generic two component $F_1,\ F_2$ hidden sector dark matter model 
discussed earlier. For definiteness
we have assumed the same $v_0$ value for $F_2$ (i.e. same $\bar m$ value) as for $A'$ 
in mirror dark matter.
We have computed the $\chi^2$ as before, minimizing over systematic uncertainties in quenching factor.
The best fit features 
$\chi^2_{min}/d.o.f$ values of 23.0/29 for the CoGeNT data set (a),
and 9.2/10 for the DAMA data set (b).
The parameter range favoured by the CoGeNT and DAMA data sets (a) and (b) discussed above is given
in figure 3 for this case.
Note that since the electric charge of $F_2$ is $\epsilon e$ rather than $Z' \epsilon e$ 
the allowed region is shifted c.f. the mirror matter case: $\epsilon \leftrightarrow \epsilon/Z'$.
We have also found that the model can fit the data for a wide range of $v_0(i)$ values:
$v_0(i) \stackrel{<}{\sim} 140\ km/s$.

The explanation of the DAMA/LIBRA and CoGeNT experiments considered here has a number of interesting features.
As noted previously\cite{foot2} 
the signals seen in these experiments arise predominately from dark matter particle interactions in
the body of their Maxwellian velocity distribution 
rather than the tail (as in the model of ref.\cite{gelmini,hooper}).
Because of this, we do not have a great deal of freedom in modifying the predicted shape
of the spectrum, and thus the agreement of the model with the spectrum observed by CoGeNT
is a non-trivial test of the theory. 
In fact in the $v_0^2(A')/v_{rot}^2 \to 0$ limit, the energy dependence of $dR/dE_R$ [Eq.(\ref{55})]
follows
exactly that of $d\sigma/dE_R$ and is proportional to $1/E_R^2$ for $v_{min} < v_{rot}$
and $d\sigma/dE_R = 0$ for $v_{min} > v_{rot}$. [Excepting here the 
energy dependence of the form factor which is relatively minor for $E_R \stackrel{<}{\sim} 1$ keVee
in germanium]. The $1/E_R^2$ dependence of $d\sigma/dE_R$ follows directly from the 
masslessness of the exchanged photon in the Feynman diagram describing the
interaction and is thus a distinctive feature of dark matter
interacting via Rutherford scattering. For finite $v_0 (A')/v_{rot}$ the $1/E_R^2$ behaviour is
expected provided that $E_R$ is sufficiently small that $v_{min} \stackrel{<}{\sim} v_{rot}$,
i.e. for
\begin{eqnarray}
E_R \stackrel{<}{\sim}  {2m_A m_{A'}^2 \over (m_A + m_{A'})^2} \ v_{rot}^2 \ .
\end{eqnarray}
For $A=Ge$, $v_{rot} = 240$ km/s and $m_{A'} \stackrel{>}{\sim} 18$ GeV (the latter suggested 
by the fit to the DAMA annual modulation signal),
we have 
\begin{eqnarray}
{dR \over dE_R} \propto {1 \over E_R^2}
\end{eqnarray}
for $E_R \stackrel{<}{\sim} 1$ keVee.
This prediction is impressively consistent with the observations as indicated in
figure 4. 
CoGeNT's spectrum falls off more rapidly than $1/E_R^2$ at $E_R \stackrel{>}{\sim} 1$ keVee.
This suggests the onset of the kinematic threshold $v_{min} \stackrel{>}{\sim} v_{rot}$
at these energies and is the origin of the $m \stackrel{<}{\sim} 30$ GeV upper limit 
indicated in figures 1,3.
\vskip 0.5cm

\centerline{\epsfig{file=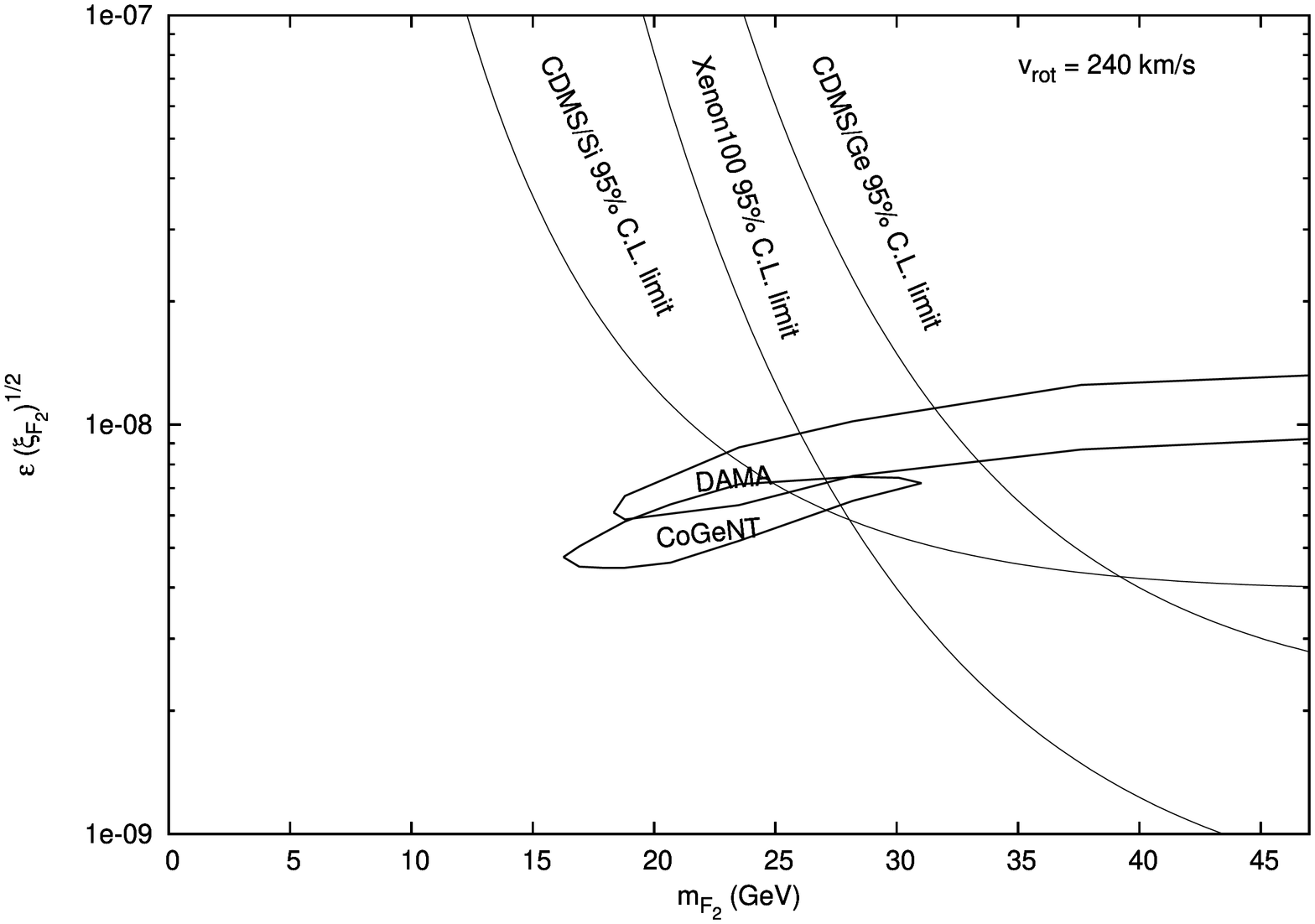,angle=270,width=15.0cm}}
\vskip 0.4cm
\noindent
{\small Figure 3: CoGeNT and DAMA $99\%$ C.L. favoured parameter ($m_{F_2},\ \epsilon \sqrt{\xi_{F_2}}$)
regions for the generic hidden sector dark matter model discussed in the text.  The reference point
$v_{rot} = 240$ km/s is assumed. Also shown are the $95\%$ exclusion curves
evaluated from the null results of the XENON100, CDMS/Si and CDMS/Ge experiments.
}

\vskip 0.8cm

The dark matter explanation of CoGeNT's spectrum offered here can be compared with the
model of ref.\cite{gelmini,hooper} which features WIMPS elastically 
scattering via a contact (four fermion) interaction rather than via Rutherford scattering.
The contact interaction produces a flat (in $E_R$) cross-section, excepting the
mild (at these energies) recoil energy dependence of the form factor.  
A rapidly falling spectrum would then only be expected if 
dark matter particles are so light that only  
particles in the
tail of the halo velocity distribution can lead to recoils with enough
energy to be observed. In such a scenario the shape of the $dR/dE_R$ spectrum
necessarily depends very sensitively on $m_{wimp}$. 
Only for $m_{wimp} \simeq 7.0$ GeV (and with standard assumptions)\cite{cogent2}
can that model account for the observed spectrum energy dependence 
$dR/dE_R \sim 1/E_R^2$ at low $E_R$.
However the energy dependence is accommodated, rather than explained, which is in contrast to
the Rutherford scattering scenarios considered here.


\vskip 0.5cm

\centerline{\epsfig{file=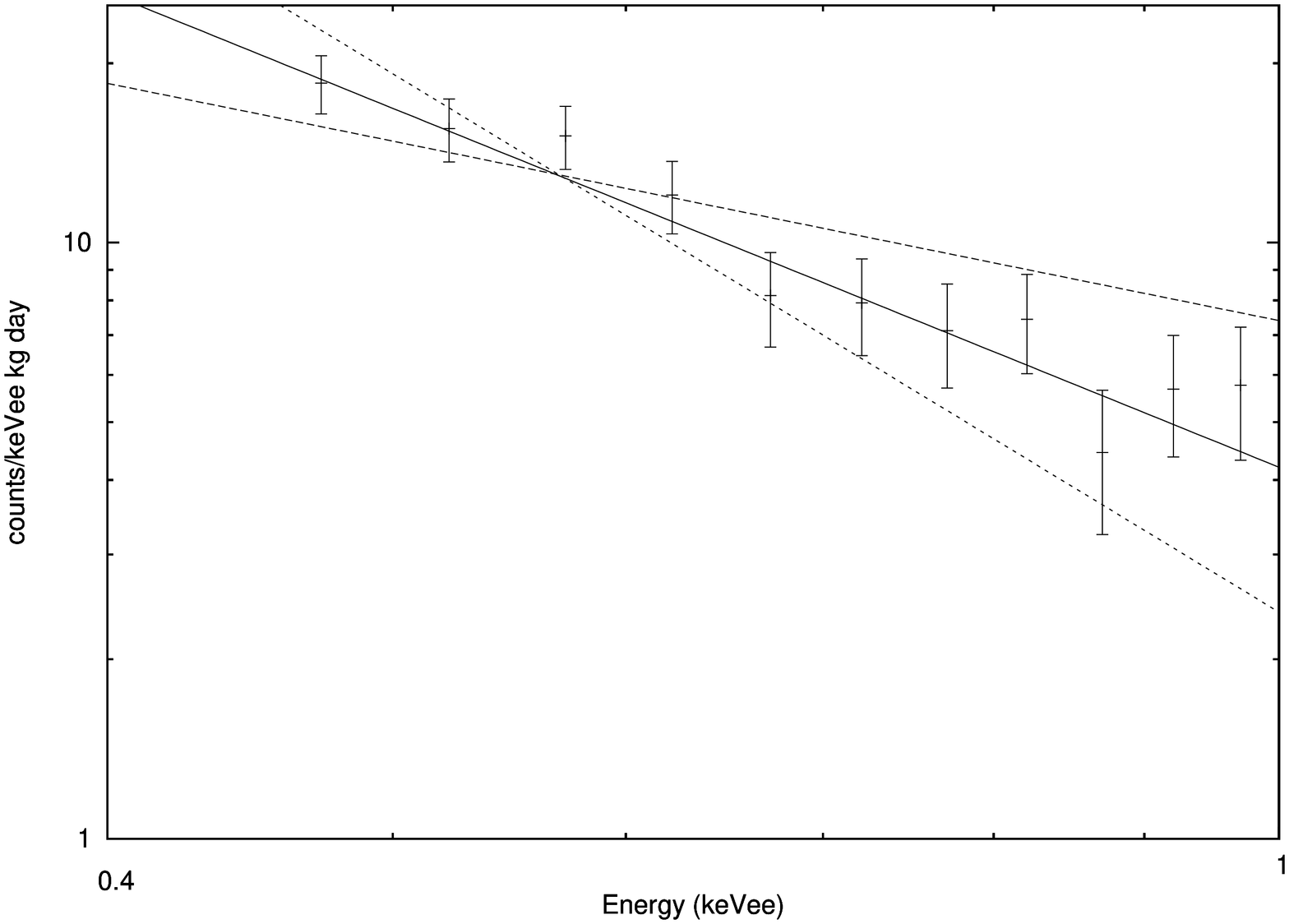,angle=270,width=15.0cm}}
\vskip 0.3cm
\noindent
{\small 
Figure 4: Low energy CoGeNT spectrum compared with $dR/dE_R \propto 1/E_R^n$ where
$n = 1$ (dashed line), $n=2$ (solid line)  and $n=3$ (dotted line).
The data clearly favour the $n = 2$ case, which is expected in the mirror dark matter/hidden
sector models considered here, and is characteristic of dark matter interacting via
elastic Rutherford scattering.
}

\vskip 0.8cm

Future experiments should be able to more clearly distinguish mirror/hidden sector models 
from other theoretical explanations - such as the one discussed in ref.\cite{gelmini,hooper} and 
many others considered in recent literature - 
by e.g. precise measurements of the annual modulation energy spectrum.
The mirror/hidden sector models predict a characteristic change in sign of the annual modulation amplitude
at low energies (see figure 6 of ref.\cite{foot2}).
Distinguishing the mirror dark matter case from the more generic hidden sector model might prove
more challenging.
Whilst the two theories give essentially identical results for the DAMA/CoGeNT experiments
mirror dark matter predicts a spectrum of particles of known masses. In particular the 
scattering of low mass components, $e'$ on electrons and $He'/H'$ on target nuclei
can ultimately be seen in very low threshold experiments. 
Higher mass sub-components, such as a $\sim Fe'$ or $\sim Ca'$ component would also be expected and
should ultimately be observed if dark matter is of the mirror type.


In conclusion, we have examined mirror/hidden sector dark matter in the light of 
CoGeNT's
more precisely measured spectrum and
annual modulation signal\cite{cogent2}.
The CoGeNT spectrum is observed to obey 
$dR/dE_R \propto 1/E_R^2$ at low energies which suggests that dark matter interacts via a
massless or light mediator (Rutherford scattering)
rather than the more commonly assumed contact (four-fermion) interaction.
Such Rutherford scattering is a feature of
mirror and more generic hidden sector dark matter models considered here and in previous
works\cite{foot1,foot2,foot2008,footold}.
We have found that such models provide an excellent fit to the data
which is easily consistent with the null results of the sensitive but higher threshold experiments, 
such as CDMS and XENON100. 
The next generation P-type Point Contact detectors, 
including CoGeNT(C-4), MAJORANA, GERDA and CDEX should be able to provide
a decisive test of these models by e.g. a precise measurement of the annual modulation energy spectrum.
These and other experiments are awaited with interest.

\vskip 1cm
\noindent
{\large Acknowledgments}

\vskip 0.2cm
\noindent
This work was supported by the Australian Research Council.


\begin{thebibliography}{999}

\bibitem{cogent}
C. E. Aalseth {\it et al.} (CoGeNT Collaboration),  Phys. Rev. Lett. 106: 131301 (2011)
[arXiv:1002.4703].

\bibitem{cogent2}
C. E. Aalseth {\it et al.} (CoGeNT Collaboration), arXiv: 1106.0650.

\bibitem{foot1}
R. Foot, Phys. Lett. B692: 65 (2010) [arXiv: 1004.1424].

\bibitem{foot2}
R. Foot, Phys. Rev. D82: 095001 (2010) [arXiv: 1008.0685].

\bibitem{hooper}
D. Hooper, J. I. Collar, J. Hall and D. McKinsey, Phys. Rev. D82: 123509 (2010) 
[arXiv: 1007.1005].

\bibitem{dm}
A. K. Drukier, K. Freese and D. N. Spergel, Phys. Rev. D33, 3495 (1986);
K. Freese, J. A. Frieman and A. Gould, Phys. Rev. D37, 3388 (1988).

\bibitem{dama1}
R. Bernabei {\it et al}. (DAMA Collaboration), 
Riv. Nuovo Cimento. 26, 1 (2003) [astro-ph/0307403]; Int. J. Mod.
Phys. D13, 2127 (2004); Phys. Lett. B480, 23 (2000).

\bibitem{dama2}
R. Bernabei {\it et al}. (DAMA Collaboration), 
Eur. Phys. J. C56: 333 (2008) [arXiv:0804.2741]; Eur. Phys. J. C67, 39 (2010) [arXiv: 1002.1028].

\bibitem{cdmsge}
Z. Ahmed {\it et al} (CDMS Collaboration), Science 327: 1619 (2010) [arXiv: 0912.3592].

\bibitem{xenon100}
E. Aprile {\it et al.} (XENON100 Collaboration), arXiv: 1104.3121.

\bibitem{footold}
R. Foot, Phys. Rev. D69, 036001 (2004) [hep-ph/0308254]; astro-ph/0403043; Mod. Phys. Lett. A19,
1841 (2004) [astro-ph/0405362]; Phys. Rev. D74, 023514 (2006) [astro-ph/0510705].

\bibitem{gelmini}
P. Gondolo and G. Gelmini, Phys. Rev. D71: 123520 (2005) [hep-ph/0504010].

\bibitem{gelmini2}
C. Savage, G. Gelmini, P. Gondolo and K. Freese, Phys. Rev. D83: 055002 (2011)
[arXiv: 1006.0972].

\bibitem{foot2008}
R. Foot, Phys. Rev. D78, 043529 (2008) [arXiv: 0804.4518].

\bibitem{he}
R. Foot and X-G. He, Phys. Lett. B267, 509 (1991). 

\bibitem{footz}
R. Foot, Phys. Lett. B699, 230 (2011) [arXiv:1011.5078].

\bibitem{feng}
J. L. Feng, M. Kaplinghat, 
H. Tu and H-B. Yu, JCAP 0907, 004 (2009) [arXiv: 0905.3039].

\bibitem{flv}
R. Foot, H. Lew and R. R. Volkas, Phys. Lett. B272, 67 (1991);
Mod. Phys. Lett. A7, 2567 (1992).

\bibitem{review}
R. Foot, Int. J. Mod. Phys. D13, 2161 (2004)
[astro-ph/0407623]; P. Ciarcelluti, Int. J. Mod. Phys. D19: 2151 (2010) [arXiv: 1102.5530].

\bibitem{some}
H. M. Hodges, Phys. Rev. D47, 456 (1993); 
Z. Berezhiani, D. Comelli and F. L. Villante,
Phys. Lett. B503, 362 (2001) [hep-ph/0008105];
L. Bento and Z. Berezhiani, Phys. Rev. Lett. 87, 231304 (2001)
[hep-ph/0107281]; 
A. Yu. Ignatiev and R. R. Volkas, Phys. Rev. D68, 023518 (2003)
[hep-ph/0304260];
R. Foot and R. R. Volkas, Phys. Rev. D68, 021304 (2003)
[hep-ph/0304261]; Phys. Rev. D69, 123510 (2004) [hep-ph/0402267];
Z. Berezhiani, P. Ciarcelluti, D. Comelli and F. L. Villante,
Int. J. Mod. Phys. D14, 107 (2005) [astro-ph/0312605];
P. Ciarcelluti, Int. J. Mod. Phys. D14, 187 (2005) [astro-ph/0409630];
Int. J. Mod. Phys. D14, 223 (2005) [astro-ph/0409633].
For pioneering work, see: S. I. Blinnikov and M. Yu. Khlopov, Sov. J. Nucl. Phys.
36, 472 (1981); Sov. Astron. 27, 371 (1983).

\bibitem{sph}
R. Foot and R. R. Volkas,
Phys. Rev. D70, 123508 (2004) [astro-ph/0407522].

\bibitem{lab1}
R. Foot, A. Yu. Ignatiev and R. R. Volkas, 
Phys. Lett. B503, 355 (2001) [arXiv: astro-ph/0011156];
R. Foot, Int. J. Mod. Phys. A19 3807 (2004) [astro-ph/0309330];
R. Foot and Z. K. Silagadze, Int. J. Mod. Phys. D14, 143 (2005) [astro-ph/0404515];
R. Foot, Phys. Lett. B699, 230 (2011) [arXiv:1011.5078].
See also, S. Davidson, S. Hannestad and G. Raffelt, JHEP 5, 3
(2000) [arXiv: hep-ph/0001179].

\bibitem{mitra}
R. Foot and S. Mitra, Astropart. Phys. 19, 739 (2003) [astro-ph/0211067]; 
Phys. Lett. A315, 178 (2003) [cond-mat/0306561];
Phys. Lett. B558, 9 (2003) [astro-ph/0301229].

\bibitem{disf}
P. Ciarcelluti and R. Foot, 
Phys. Lett. B679, 278 (2009) [arXiv: 0809.4438].

\bibitem{helm}
R. H, Helm, Phys. Rev. 104, 1466 (1956). 

\bibitem{smith}
J. D. Lewin and P. F. Smith, Astropart. Phys. 6, 87 (1996).




\bibitem{newstudy}
N. Bozorgnia, G.B. Gelmini and P. Gondolo, JCAP 1011:019 (2010) [arXiv: 1006.3110];
JCAP 1011:028 (2010) [arXiv: 1008.3676].

\bibitem{bbn}
P. Ciarcelluti and R. Foot, Phys. Lett. B690, 462 (2010) [arXiv:1003.0880].


\bibitem{reidrecent}
A. Brunthaler {\it et al}. arXiv: 1102.5350.




\bibitem{footescattering}
R. Foot, Phys. Rev. D80, 091701 (2009) [arXiv: 0909.3126].

\bibitem{cdmssi}
J. P. Filippini, Ph.D thesis, 2008.


\bibitem{xenonlow}
J. Angle {\it et al}. (XENON10 Collaboration), arXiv: 1104.3088.

\bibitem{cdmslow}
Z. Ahmed {\it et al}. (CDMS Collaboration), Phys. Rev. Lett. 106: 131302 (2011) 
[arXiv: 1011.2482].

\bibitem{collarlow}
J. I. Collar, arXiv: 1010.5187; 1103.3481.

\bibitem{mm3}
R. Foot, Phys. Rev. D81, 087302 (2010) [arXiv:1001.0096].




\end{thebibliography}
\end{document}